\documentclass[useAMS,usenatbib]{mn2e}
\usepackage{txfonts}
\usepackage{natbib}
\usepackage[all]{xy}
\usepackage[dvips]{graphicx}
\usepackage{array}

\usepackage{hyperref}
\usepackage[usenames,dvipsnames]{color}
\definecolor{menublue}{rgb}{0.0,0.0,0.5}
\definecolor{citegreen}{rgb}{0.0,1.0,0.0}
\definecolor{urlred}{rgb}{1.0,0.0,0.0}
\hypersetup{bookmarksopen,pdfstartview={FitH},colorlinks=true, breaklinks=true,menucolor=menublue,urlcolor=urlred,citecolor=citegreen,linkcolor=blue}
\usepackage[all]{hypcap}

\def\del#1{{}}

\newcommand{\ltsima}{$\; \buildrel < \over \sim \;$}
\newcommand{\lsim}{\lower.5ex\hbox{\ltsima}}
\newcommand{\gtsima}{$\; \buildrel > \over \sim \;$}
\newcommand{\gsim}{\lower.5ex\hbox{\gtsima}}
\newcommand{\bra}{\langle}
\newcommand{\ket}{\rangle}

\newcommand{\dd}{\mathrm{d}}

\newcommand{\ci}{\mathrm{i}}
\newcommand{\vecx}{\bmath{x}}

\newcommand{\trace}{\mathrm{tr}}

\newcommand{\chip}{{\chi^\prime}}

\newcommand{\lprime}{\ell^\prime}

\newcommand{\lhat}{\hat{L}}
\newcommand{\vect}{\bmath{\theta}}
\newcommand{\veca}{\bmath{\alpha}}

\title[lensing of aligned galaxies]
{Weak gravitational lensing of intrinsically aligned galaxies}
\author[A. Giahi-Saravani, B.M. Sch{\"a}fer]
{Aram Giahi-Saravani\thanks{e-mail: aram@ari.uni-heidelberg.de} and Bj{\"o}rn Malte Sch\"afer\\
Zentrum f{\"u}r Astronomie der Universit{\"a}t Heidelberg, Philosophenweg 12, 69120 Heidelberg, Germany}

\begin{document}
\pagerange{\pageref{firstpage}--\pageref{lastpage}}
\pubyear{2013}
\maketitle
\label{firstpage}

\begin{abstract}
Subject of this paper is the weak lensing effect on galaxies that show intrinsically correlated ellipticities. In our model, we investigate the distortion of the ellipticity field if the galaxies experience an apparent shift in their position by weak lensing deflection and compare this effect to the shearing effect induced by tidal fields. Starting with a derivation of intrinsic ellipticity spectra by employing a tidal torquing model generating galactic angular momenta, we model the galaxy ellipticity by assuming that the galactic disk forms perpendicularly to the host halo angular momentum direction and derive intrinsic ellipticity $E$-mode and $B$-mode spectra from the angular momentum statistics. The lensing effect on the ellipticity field is modeled by employing the methodology developed in the framework of lensing of the cosmic microwave background polarisation. For EUCLID, ellipticity correlations are altered by lensing deflection on multipoles $\ell\gsim1000$ by $\sim5\%$ for the ellipticity $E$-modes and by $\sim30\%$ for the $B$-modes, while a shallower survey would exhibit larger changes on larger angular scales. In addition to the convolving effect of lensing on the ellipticity spectra we investigate the $E$/$B$-mode conversion, and discuss the possibility of measuring correlations between different multipoles which is evoked by the homogeneity breaking effect of the lensing displacement. Our conclusion is that although shape correlations generated by weak gravitational shear is dominant, the shifting effect due to lensing is shaping the ellipticity spectra on small angular scales and causes a number of interesting phenomena, which might be observable by future surveys.
\end{abstract}

\begin{keywords}
cosmology: large-scale structure, gravitational lensing, methods: analytical
\end{keywords}

\section{Introduction}
Weak gravitational lensing by the cosmic-large scale structure provides a measurement of the cosmic tidal field and provides sensitivity on cosmological parameters due its dependence on the geometry of the cosmological model and the growth rate of matter perturbations \citep[for reviews, see][]{2001PhR...340..291B, 2008ARNPS..58...99H, 2010CQGra..27w3001B}. Weak lensing data taken by future surveys such as EUCLID, DES and LSST has the potential of putting tight constraints on cosmological parameters and to distinguish between dark energy models and those with modified gravity \citep[for a comprehensive summary, see][]{2012arXiv1201.2434W}.

In analysing weak lensing data it is usually assumed that weak lensing is the only effect causing correlations in the shapes of galaxies, which is sadly not the case as correlations in the shapes of neighbouring galaxies are naturally explained by their correlated angular momenta due to similarities in their formation processes, most importantly the tidal shearing experienced by their host haloes \citep[for a review on angular momentum models and intrinsic alignments, see][]{2009IJMPD..18..173S}.

Angular momentum is introduced into haloes in their formation as a consequence of tidal shearing which  has been investigated using perturbation theory, including the deformation of forming haloes, as well as by numerical studies \citep{1995MNRAS.276..115C, 1996MNRAS.282..455C, 1996MNRAS.282..436C, 2001MNRAS.320L...7C, 2001MNRAS.323..713C,2007arXiv0709.1106L, 2012arXiv1201.5794C}. There, the role of tidal torquing in the angular momentum build-up during galaxy formation is supported, but indicate that tidal torquing models might be predicting too high values for the amount of angular momentum \citep{2001ApJ...555..240B, 2001MNRAS.323..713C, 2002MNRAS.332..325P, 2002MNRAS.332..339P, 2007MNRAS.381...41H} while perturbation theory seems to yield fairly reliable results for the angular momentum direction and explaining alignment effects of the halo with the ambient large-scale tidal field.

Angular momentum alignments give rise to a correlation in intrinsic ellipticities between neighbouring galaxies if the angular momentum direction of the galaxy corresponds to the one of the host halo. The scale of this correlation is predicted to be in the range of $\sim1~\mathrm{Mpc}/h$~\citep{2001ApJ...559..552C, 2001PASA...18..198N, 2012MNRAS.421.2751S}. Hence, intrinsic alignments influence the angular momentum correlation on small scales and can be of significance for high precision observations such as the future surveys mentioned above. Estimates of intrinsic alignment spectra arising from angular momentum based-models can be found in the works of~\citet{2000ApJ...545..561C, 2001ApJ...559..552C, 2002ApJ...568...20C, 2002MNRAS.332..788M}. Due to their large complexity, \citet{2010MNRAS.402.2127S} devised a model which allows a much easier handling of intrinsic alignments.

Direct identification of the symmetry axis of the disc with the host halo angular momentum would lead to overestimation of the ellipticity alignments and is flawed as there might  be large misalignments between these two vectors due to baryonic physics \citep{2005ApJ...627L..17B, 2013arXiv1301.3143S}. Thus, we want to think of our ellipticity spectra as upper limits. Whereas the alignment model for spiral galaxies a quadratic dependency of the ellipticity on the tidal field is assumed \citet{2001ApJ...559..552C}, elliptical galaxies follow a simpler linear alignment mechanism \citep{2004PhRvD..70f3526H}. Sadly, the processes determining the orientation of the stellar disk inside a dark matter halo are not amenable to analytic calculations and even numerical simulations struggle to reach consensus about properties of galactic disks in a cosmological volume.

Intrinsic alignments are difficult to quantify \citep{2012ApJ...747....7G} but detections have been reported in a number of data sets, for instance in the Tully-catalogue \citep{2000ApJ...543L.107P}, the Point Source Catalogue Redshift survey \citep{2002ApJ...567L.111L}, in the Sloan Digital Sky survey \citep{2004MNRAS.353..529H, 2006MNRAS.367..611M, 2007ApJ...670L...1L, 2009ApJ...694L..83O, 2011A&A...527A..26J}, in combination with the 2-degree Field Galaxy Redshift survey \citep{2007MNRAS.381.1197H} and the WiggleZ Dark Energy Survey \citep{2011MNRAS.410..844M}, which allowed tests of alignment models \citep{2011JCAP...05..010B} and the determination of alignment model parameters. \citet{2011MNRAS.410..844M} in particular provide a detailed description of the technical difficulties in measuring intrinsic correlations at redshifts relevant for gravitational lensing, and give fits for the intrinsic ellipticity correlation and the ellipticity-density cross-correlation. Some studies, however, remain sceptical about these detection reports \citep{2011MNRAS.tmp.1665A}. 
 
As observations of galaxy ellipticities become increasingly precise, the study intrinsic alignments of galaxies based on angular momentum models will gain much interest in the future. Such models \citep{2000ApJ...545..561C, 2001ApJ...559..552C, 2002ApJ...568...20C, 2002MNRAS.332..788M} have been applied to estimate the significance of intrinsic alignments in the weak lensing convergence spectrum \citep{2000MNRAS.319..649H, 2003MNRAS.339..711H, 2004MNRAS.347..895H} and bispectrum  \citep{2008arXiv0802.3978S}. Techniques for separating weak lensing data from intrinsic alignments range from discarding close galaxy pairs \citep{2002A&A...396..411K,2003A&A...398...23K} to introducing a weighting in order to null out their contribution \citep{2008arXiv0804.2292J, 2011MNRAS.410.1677K}, amplifying them relative to the weak lensing-induced correlations \citep{2003A&A...398...23K, 2010arXiv1009.2024J, 2010A&A...517A...4J, 2010A&A...523A...1J} or to use the non-zero vortical modes ($B$-modes) of the intrinsic ellipticity field \citep{2002ApJ...568...20C}. With a model for intrinsic alignments at hand, it is of course possible to fit both the weak lensing ellipticity correlations and the intrinsic correlations at the same time, possibly constraining intrinsic alignment parameters \citep{2009ApJ...695..652B, 2011arXiv1109.4536K} or use self-calibration techniques \citep{2010ApJ...720.1090Z}. Many of the removal techniques require very good control of redshift estimates \citep{2007NJPh....9..444B} and can be extended to deal with cross-correlations between weak lensing and intrinsic alignments \citep{2005A&A...441...47K}. A puzzling result is that the contaminating effect of intrinsic alignments predicted from different angular momentum models in weak lensing data can be dramatically different. If intrinsic alignments are present in weak lensing data but not included in the model for the data when determining cosmological parameters, biases in the parameter estimates have to occur: Whereas \citet{2010MNRAS.408.1502K, 2012MNRAS.424.1647K} using a linear model found most biases in the dark energy equation of state, \citet{2012arXiv1207.5939C} arrive at the conclusion that biases arise mostly in $\Omega_m$ and $\sigma_8$.

Motivation for this paper is to study the influence of weak gravitational lensing on the observed ellipticity correlation, arising from the lensing effect of the intervening matter distribution. In order to assess the significance of intrinsic alignments in weak lensing measurements, \citet{2001ApJ...559..552C} found that the intrinsic signal is between 1 to 10 per cent of the measured lensing signal for a deep reaching survey. Here, our aim is to investigate the implications of the distortion (caused by weak gravitational shear) of galaxy ellipticities and the shift of galaxy positions by weak lensing on the shape of the $E$- and $B$-mode spectra of the lensed ellipticity correlation function. We will thus employ a formalism based on lensing of the cosmic microwave background (CMB) polarization~\citep{1996ApJ...463....1S, 2001ApJ...554...67H,2005PhRvD..71j3010C,2006PhR...429....1L}. It should be mentioned that in contrast to the lensing of the CMB where only the shifting effect is of importance and the polarization of the CMB photons remains unchanged, galaxy lensing involves also a distortion of ellipticities and a change of their orientation. In this regard, the analogy between the polarization of the CMB and the orientation of galaxy ellipticities has to be viewed differently. We use the model of  \citet{2001ApJ...559..552C} for deriving a correlation function for the intrinsic ellipticities and apply the CMB formalism to it in order to observe characteristic features of weak lensing, such as the mixing of E- and B-modes. We expect a suppression of ellipticity correlations due to the shifting effect which dilutes and randomises the galaxy position at which ellipticities are measured.

In his paper we collect necessary results from cosmology, angular momentum generation in galaxies, ellipticity correlations, and weak lensing in Sect.~\ref{sect_cosmology}, before describing in detail the formalism we use for computing lensed ellipticity spectra in Sect.~\ref{sect_evolution}, along with an investigation of the effects predicted by our formalisms for the case of a high- and a low-redshift galaxy sample. We summarise our main results in Sect.~\ref{sect_summary}. As reference model we work with a basic spatially flat $w$CDM model with Gaussian adiabatic initial perturbations in cosmic density field. Model parameters were set to be $\Omega_m = 0.25$, $n_s = 1$, $\sigma_8 = 0.85$, $\Omega_b=0.04$ and the Hubble-radius $c/H_0 = 2996.9~\mathrm{Gpc/h}$, with $h=0.72$. The dark energy equation of state parameter $w$ was assumed to be constant at a value of $-0.9$.

\section{cosmology}\label{sect_cosmology}

\subsection{Dark energy cosmologies}
The time evolution of isotropic Friedmann-universe with homogeneous dark matter and dark energy is described by the Hubble function $H(a)=\dd\ln a/\dd t$, which is given by
\begin{equation}
\frac{H^2(a)}{H_0^2} = \frac{\Omega_m}{a^{3}} + (1-\Omega_m)\exp\left(3\int_a^1\dd\ln a\:(1+w(a))\right),
\end{equation}
with the matter density parameter $\Omega_m$ and the dark energy equation of state function $w(a)$. Spatial flatness requires the dark energy density to be $1-\Omega_m$. The comoving distance $\chi$ can be computed from the scale factor $a$,
\begin{equation}
\chi = c\int_a^1\:\frac{\dd a}{a^2 H(a)}.
\end{equation}
For the galaxy redshift distribution $n(z)\dd z$, we use a standard shape
\begin{equation}
n(z) = n_0\left(\frac{z}{z_0}\right)^2\exp\left(-\left(\frac{z}{z_0}\right)^\beta\right)\dd z
\quad\mathrm{with}\quad \frac{1}{n_0}=\frac{z_0}{\beta}\Gamma\left(\frac{3}{\beta}\right),
\end{equation}
with $\beta=3/2$. We choose $z_0$ such that the distribution has a median redshift of 0.9 corresponding to EUCLID \citep{2007MNRAS.381.1018A} which we contrast with a galaxy distribution of identical shape but with a much lower median of 0.3. We will refer to the two application cases as the high and low redshift galaxy sample, respectively. The distribution can be rewritten in terms of comoving distance using the relation $p(z)\dd z = p(\chi)\dd\chi$ with $\dd z/\dd\chi = H(\chi)/c$.

\subsection{CDM power spectrum}
The statistical properties of the overdensity field $\delta$ is in the case of homogeneous, isotropic and Gaussian fluctuations described by the spectrum $P_\delta(k)$. Inflationary models suggest the ansatz
\begin{equation}
P_\delta(k)\propto k^{n_s}T^2(k),
\end{equation}
with the transfer function $T(k)$. This function is approximated by
\begin{equation}
T(q) = \frac{\ln(1+2.34q)}{2.34q}M(q)^{-\frac{1}{4}},
\label{eqn_cdm_transfer}
\end{equation}
\citep[see][]{1986ApJ...304...15B}, with the polynomial
\begin{equation}
M(q) = 1+3.89q+(16.1q)^2+(5.46q)^3+(6.71q)^4
\end{equation}
The transfer function depends most strongly on the parameters $\Omega_m$ and $h$ which form the shape-parameter $\Gamma$,
\begin{equation}
\Gamma = \Omega_m h\:\exp\left(-\Omega_b\left[1+\frac{\sqrt{2h}}{\Omega_m}\right]\right),
\end{equation}
and shows slight corrections due to the baryon density $\Omega_b$ \citet{1995ApJS..100..281S}. With $\Gamma$, the wave vector $k$ is scaled according to $q = k/\Gamma$. The linearly evolved spectrum $P(k)$ is normalised to show the variance $\sigma_8^2$ on a comoving scale of $R=8~\mathrm{Mpc}/h$,
\begin{equation}
\sigma^2_R = \int\frac{k^2\dd k}{2\pi^2}\: P_\delta(k) W^2(kR)
\end{equation}
with a Fourier transformed spherical top hat filter function, $W(x)=3j_1(x)/x$. $j_\ell(x)$ is the spherical Bessel function of the first kind of order $\ell$ \citep{1972hmf..book.....A}. Weak lensing spectra will be computed for both linear and nonlinear structures, for the latter we employ the extension of $P(k)$ to nonlinear structure formation proposed by \citet{2003MNRAS.341.1311S}. In the limit of small perturbations $\left|\delta\right|\ll 1$, cosmic structure formation is linear and homogeneous, $\delta(\bmath{x},a) = D_+(a)\delta(\bmath{x},a=1)$. The time evolution of the cosmic density field is then given by the growth function $D_+(a)$, that follows from the growth equation \citep{1997PhRvD..56.4439T, 1998ApJ...508..483W, 2003MNRAS.346..573L},
\begin{equation}
\frac{\dd^2}{\dd a^2}D_+(a) + \frac{1}{a}\left(3+\frac{\dd\ln H}{\dd\ln a}\right)\frac{\dd}{\dd a}D_+(a) = 
\frac{3}{2a^2}\Omega_m(a) D_+(a),
\label{eqn_growth}
\end{equation}
such that the spectrum $P(k)\propto D_+^2(a)$, while the nonlinear extension to $P(k)$ has its own parametrised time evolution based on $\Omega_m(a)$, 
\begin{equation}
\frac{\Omega_m(a)}{\Omega_m} = \frac{1}{a^3}\frac{H_0^2}{H^2(a)},
\end{equation}
which can be derived from the adiabaticity of the cosmic fluids.

\subsection{Angular momentum from tidal shearing}
Angular momentum generation in CDM-haloes is a Lagrangian perturbative process \citep{1949MNRAS.109..365H, 1955MNRAS.115....2S, 1969ApJ...155..393P, 1970Afz.....6..581D, 1984ApJ...286...38W}: The variation of velocities displacing a protogalactic region acts as a torque which generates angular momentum $L_\alpha$:
\begin{equation}
L_\alpha = a^3 H(a)\frac{\dd D_+}{\dd a}\epsilon_{\alpha\beta\gamma} I_{\beta\delta}\Phi_{,\delta\gamma},
\end{equation}
with the tidal shear being defined as the Hessian of the gravitational potential and the inertia tensor measuring the second moments of the mass distribution inside the protohalo,
\begin{eqnarray}
\Phi_{,\delta\gamma} & \equiv & \frac{\partial^2\Phi}{\partial x_\delta\partial x_\gamma}
\quad\mathrm{and}\\
I_{\beta\delta} & \equiv & 
\Omega_m\rho_\mathrm{crit}
\int\dd^3q\:\delta(\bmath{q})(\bmath{q} - \bar{\bmath{q}})_\beta(\bmath{q} - \bar{\bmath{q}})_\delta,
\end{eqnarray}
respectively, with implicit summation over repeated indices. $\bmath{q}$ are Lagrangian coordinates moving along with the halo's centre of gravity $\bar{\bmath{q}}$. Note that the second moments of the mass distribution $I_{\beta\delta}$ is referred to as the inertia tensor despite the fact that it has a different ordering of the axes.

This relation reflects the interesting requirement of misalignments between the shear and inertia eigensystems which is necessary for angular momentum generation: Only the antisymmetric tensor $X^-_{\beta\gamma} = \sum_\delta (I_{\beta\delta}\Phi_{,\delta\gamma} - \Phi_{,\beta\delta}I_{\delta\gamma})/2$ is relevant for the angular momentum \citep{2012MNRAS.421.2751S}, $L_\alpha\propto X^-_{\beta\gamma}$, because the contraction of the symmetric contribution $X^+_{\beta\gamma}$ with the antisymmetric $\epsilon_{\alpha\beta\delta}$ vanishes. The antisymmetric tensor $X_-$ is equal to the commutator $[I_{\beta\delta}, \Phi_{,\delta\gamma}]$ which suggests that the tidal shear and the inertia are not allowed to be simultaneously diagonalisable and must not have a common eigensystem, otherwise angular momentum can not arise.

In this work we use the angular momentum-based ellipticity correlation model proposed by \citet{2001ApJ...559..552C}. There, ellipticities are set into relation to tidal shear by means of a conditional probability distribution $p(\bmath{L}|\Phi_{,\alpha\beta})\dd\bmath{L}$. Such a distribution has been proposed by \citet{2001ApJ...555..106L} as being Gaussian with the covariance
\begin{equation}
\mathrm{cov}(L)_{\alpha\beta} \propto \frac{1}{3}\left(\frac{1+a}{3}\delta_{\alpha\beta} - a\: (\hat\Phi^2)_{\alpha\beta}\right),
\end{equation}
which acquires a dependence on the tidal shear tensor, that has been normalised and made trace-free, $\trace(\hat\Phi) = 0$ and $\trace(\hat\Phi^2) = 1$. In this way, the variance of the angular momentum field varies with the tidal shear, and the randomness of the angular momentum field is controlled by the misalignment parameter $a$, which describes the average orientation of the protohalo's inertia to the tidal shear eigensystem. $a$ has been measured in numerical simulation to be the value $\simeq0.25$, and we will use this value in this work. 

This description is valid on scales where the correlations between inertia tensors are negligible and is sufficient because we are only interested in the angular momentum direction, as will be explained in the next section. Therefore, one does not need the variance $\bra\bmath{L}^2\ket$ of the angular momentum field as a parameter and it is possible to marginalise the distribution over the magnitude $L$,
\begin{equation}
p(\lhat|\Phi_{,\alpha\beta}) = \int L^2\dd L\: p(\bmath{L}|\Phi_{,\alpha\beta}).
\end{equation}
In this picture correlations between angular momenta can be traced back to correlations between tidal shears that neighbouring galaxies experienced in building up their angular momenta. It should be noted, however, that recent investigations provide evidence that the nonlinear evolution of galactic angular momenta can be vorticity driven \citep{2012arXiv1212.1454L, 2013arXiv1301.0348L} as an alternative to tidal torquing.

\subsection{Intrinsic ellipticity correlations}
We assume that the halo angular momentum axis and the symmetry axis of the galactic disk are parallel. The tilting of the disk relative to the line of sight determines the aspect ratio under which the galactic disk is viewed and therefore the galaxy's ellipticity. In this way, the angular momentum direction $\lhat = \bmath{L}/L$ with $L = \left|\bmath{L}\right|$ determines the ellipticity $\epsilon$. In this picture, ellipticity correlations are derived from angular momentum direction correlations and ultimately from tidal shear correlations \citep{2000MNRAS.319..649H, 2001ApJ...559..552C, 2002ApJ...568...20C, 2002MNRAS.332..788M, 2003MNRAS.339..711H}. Specifically, the two components of the ellipticity field can be combined to form the complex ellipticity with $\epsilon_+$ as the real and $\epsilon_\times$ as the complex part, and related to the angular momentum direction $\lhat$:
\begin{equation}
\epsilon=\epsilon_+ +\ci\epsilon_\times
\quad\mathrm{with}\quad
\epsilon_+ = \alpha\frac{\lhat_x^2-\lhat_y^2}{1+\lhat_z^2},\quad
\epsilon_\times = 2\alpha\frac{\lhat_x\lhat_y}{1+\lhat_z^2},
\end{equation}
when the coordinate system is aligned with its $z$-axis parallel to the line of sight. Under a rotation of the coordinate frame by an angle $\varphi$ the complex ellipticity transforms according to the relation $\epsilon\rightarrow\exp(2\ci\varphi)\epsilon$, in accordance with the spin-2 property of the ellipticity field. $\alpha$ is the disk thickness parameter and can be used for describing a weaker dependence of $\epsilon$ on $\lhat$ if the galactic disk has a finite thickness. We use the numerical value $\alpha = 0.75$ which has been measured in the APM galaxy sample by \citet{2001ApJ...559..552C}.

Linking the ellipticity components $\epsilon_+$ and $\epsilon_\times$ to the angular momentum direction $\lhat$ and ultimately to the tidal shear tensor by means of the conditional probability density $p(\bmath{L}|\Phi_{,\alpha\beta})\dd\bmath{L}$ provides means to derive ellipticity correlations in terms of tidal shear correlations, and ultimately as functions of the CDM-spectrum. Because we are interested in ellipticity correlations of galaxies, we smooth the CDM-spectrum with a Gaussian filter on the mass-scale  $3\times10^{11}M_\odot/h$. In this way, one can write down a 3-dimensional correlation function of the ellipticity field as a function of moments $\zeta_n(r)$ \citep[see][]{2001ApJ...559..552C} of the CDM-spectrum and finally to carry out a Limber projection for obtaining the angular correlation function of the ellipticity field. The correlation function can be Fourier-transformed to yield the $E$-mode and $B$-mode ellipticity spectra (c.f. Sect.~\ref{sect_spin2}). The related model proposed by \citet{2002MNRAS.332..788M} yields very similar shapes for the ellipticity spectra but predicts smaller amplitudes.

\subsection{Weak gravitational lensing}
The lensing potential $\psi$ is defined as the line of sight projected gravitational potential $\Phi$
\begin{equation}
\psi = \int_0^{\chi_H}\dd\chi\: \frac{G(\chi)}{\chi}\:\Phi,
\end{equation}
with the lensing-efficiency weighted galaxy distribution $G(\chi)$ as a weighting function,
\begin{equation}
G(\chi) = \int_\chi^{\chi_H}\dd\chip\:n(\chip)\left(1-\frac{\chi}{\chip}\right)
\end{equation}
The lensing deflection angle $\alpha$ is obtained from the lensing potential $\psi$ by differentiation, $\alpha_i = \partial_i\psi$. Further differentiation yields the projected tidal field $\bpsi\equiv\partial_i\partial_j\psi$ which can be decomposed with the Pauli-matrices $\sigma_\alpha$, because they constitute a basis for the vector space of $2\times2$ matrices,
\begin{equation}
\bpsi
= \sum_{\alpha=0}^3\: a_\alpha\sigma_\alpha 
= (1-\kappa)\sigma_0 - \gamma_+\sigma_1 - \gamma_\times\sigma_3,
\end{equation}
\citep[c.f.][]{1972hmf..book.....A, 2005mmp..book.....A}, where $\sigma_0$ denotes the $2$-dimensional unit matrix. For $\alpha=1,2,3$, the Pauli-matrices have the properties $\sigma_\alpha^2=\sigma_0$ and $\trace(\sigma_\alpha)=0$. Due to the property $\sigma_\alpha\sigma_\beta = \sigma_0\delta_{\alpha\beta} + \ci\epsilon_{\alpha\beta\gamma}\sigma_\gamma$ of the Pauli-matrices, the coefficients $a_\alpha$ can be recovered by using $a_\alpha = \frac{1}{2}\trace(\bpsi\sigma_\alpha)$. In particular, one identifies the weak lensing convergence $\kappa = \frac{1}{2}\trace(\bpsi\sigma_0)$ with the unit matrix $\sigma_0$ and the two components of shear $\gamma_+=\frac{1}{2}\trace(\bpsi\sigma_1)$ and $\gamma_\times=\frac{1}{2}\trace(\bpsi\sigma_3)$. The standard expression for $\kappa$ can be recovered with $\kappa = \frac{1}{2}\trace(\bpsi\sigma_0) = \frac{1}{2}\sum_i\partial_i\partial_i\psi = \frac{1}{2}\Delta_\theta\psi = \frac{1}{2}\mathrm{div}_\theta\veca$ with the deflection angle $\veca = \nabla_\theta\psi$. 

The two components of lensing shear are combined to form the complex shear $\gamma = \gamma_++\ci\gamma_\times$ with the transformation property $\gamma\rightarrow\gamma\exp(2\ci\varphi)$ under a rotation of the coordinate frame by an angle $\varphi$. Violations of the symmetry of $\bpsi$ are very small and might e.g. be caused by geodesic effects such as lens-lens-coupling and Born-corrections \citep[see][for a detailed computation]{2006JCAP...03..007S,2002ApJ...574...19C,2009arXiv0910.3786K, 2010PhRvD..81h3002B}. Therefore, an expansion coefficient for the contribution due to $\sigma_2$ which parametrises image rotation was neglected. In the limit of weak lensing, the galaxy ellipticity is transformed according to $\epsilon\rightarrow\epsilon+\gamma$.

\subsection{Spectrum of the weak lensing potential}
The spectrum of the weak lensing potential follows from substituting the line of sight-expression $\psi = \int\dd\chi\: G(\chi)/\chi\:\Phi$ into the Limber-equation \citep{1954ApJ...119..655L},
\begin{equation}
C_\psi(\ell) = \int_0^{\chi_H}\frac{\dd\chi}{\chi^4}\:G(\chi)^2\:P_\Phi(k = \ell/\chi,a(\chi)),
\end{equation}
where we used the most basic flat-sky description. The power spectrum $P_\Phi(k,a)$ of the gravitational potential $\Phi$ at the scale-factor $a$ follows from the comoving Poisson equation $\Delta\Phi=3H_0^2\Omega_m/(2a)\delta$ and is related to the density power spectrum $P_\delta(k,a)$ by
\begin{equation}
P_\Phi(k,a) = \left(\frac{3\Omega_m}{2}\frac{D_+(a)}{a}\right)^2\frac{P_{\delta}(k)}{(\chi_Hk)^4},
\end{equation}
with the Hubble distance $\chi_H=c/H_0$ making the $k^{-4}$-factor dimensionless. By differentiation one obtains the spectrum $C_\alpha(\ell) = \ell^2C_\psi(\ell)$ of the deflection angle $\veca = \nabla_\theta\psi$ and the spectrum $C_\kappa(\ell) = \ell^4/4\: C_\psi(\ell)$ of the weak lensing convergence $\kappa = \frac{1}{2}\mathrm{div}_\theta\veca = \frac{1}{2}\Delta_\theta\psi$, which is equal to the $E$-mode correlation function $C^\gamma_E(\ell)$ of the weak lensing shear $\gamma$ in the absence of $C^\gamma_B(\ell)$ (c.f. the following section).

\begin{figure}
\begin{center}
\resizebox{\hsize}{!}{\includegraphics{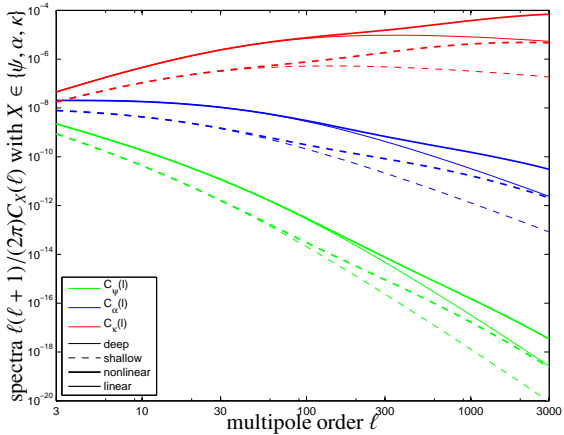}}
\end{center}
\caption{Angular spectrum $C_\psi(\ell)$ (green line) of the lensing potential $\psi$, the spectrum $C_\alpha(\ell)\equiv \ell^2C_\psi(\ell)$ (blue line) of the lensing deflection field $\alpha=\nabla_\theta\psi$ and the spectrum $C^\gamma_E(\ell) = C_\kappa(\ell) = \ell^4/4\:C_\psi(\ell)$ of the $E$-mode shear, for the high redshift galaxy sample (solid lines) and the low redshift galaxy sample (dashed lines), comparing the prediction for a linear CDM spectrum (thin lines) with a nonlinear one (thick lines).}
\label{fig_spectrum}
\end{figure}

The angular spectrum $C_\psi(\ell)$ of the weak lensing potential $\psi$ resulting from the Limber-projection of $P_\Phi(k)$ is depicted in Fig.~\ref{fig_spectrum} along with the spectrum $C_\alpha(\ell)$ of the lensing deflection angle $\veca$ and the weak lensing shear spectrum $C^\gamma_E(\ell)$, for the high redshift galaxy sample with $z_\mathrm{med}=0.9$ and the low redshift galaxy sample with $z_\mathrm{med}=0.3$. Furthermore, predictions for the spectra using a linear and a nonlinear CDM-spectrum $P(k)$ are contrasted.

\subsection{Correlations of spin-2 fields}\label{sect_spin2}
Both the galaxy ellipticities and the Stokes-parameters of the CMB-polarisation form a tensorial spin-2 field, which means that rotations of the coordinate frame by an angle $\varphi$ give rise to a transformation of the tensor components as $\epsilon\rightarrow\exp(2\ci\varphi)\epsilon$ and $P\rightarrow \exp(2\ci\varphi) P$, when the ellipticity is written as a complex ellipticity $\epsilon = \epsilon_++\ci\epsilon_\times$ and the polarisation tensor $P$ is composed of the Stokes parameters $Q$ and $U$ according to $P = U+\ci Q$.

Correlations between two points $\vect_1$ and $\vect_2$ separated by the distance $\theta$ of a spin-2 field such as the complex polarisation $P=Q+\ci U$, the complex ellipticity $\epsilon = \epsilon_++\ci\epsilon_\times$ or the weak lensing shear $\gamma = \gamma_++\ci\gamma_\times$ are described in terms of two functions $\xi_\pm(\theta)$,
\begin{eqnarray}
\xi_+(\theta) & = & 
\bra \epsilon_+(\vect_1)\epsilon_+(\vect_2)\ket + 
\bra\epsilon_\times(\vect_1)\epsilon_\times(\vect_2)\ket, \\
\xi_-(\theta) & = & 
\bra \epsilon_+(\vect_1)\epsilon_+(\vect_2)\ket - 
\bra\epsilon_\times(\vect_1)\epsilon_\times(\vect_2)\ket.
\end{eqnarray}
The correlation functions $\xi_\pm(\theta)$ are constructed from the variances of the components $\epsilon_+$ and $\epsilon_\times$ using vanishing cross-correlations, $\bra\epsilon_+\epsilon_\times\ket=0$. They can be transformed to the spectra $C_E^\epsilon(\ell)$ and $C_B^\epsilon(\ell)$ of the gradient ($E$) and vorticity ($B$) modes of the ellipticity field,
\begin{eqnarray}
C_E^\epsilon(\ell) & = & \pi\int\theta\dd\theta 
\left[\xi_+(\theta)J_0(\ell\theta) + \xi_-(\theta)J_4(\ell\theta)\right],
\label{eqn_e_transform}\\
C_B^\epsilon(\ell) & = & \pi\int\theta\dd\theta 
\left[\xi_+(\theta)J_0(\ell\theta) - \xi_-(\theta)J_4(\ell\theta)\right],
\label{eqn_b_transform}
\end{eqnarray}
by Fourier transform \citep{1992ApJ...388..272K, 2002A&A...389..729S, 2007A&A...462..841S,2010MNRAS.401.1264F}. Completely analogous formulae apply for the description of the angular correlation properties of the weak lensing shear and their transformation to Fourier space yielding $C^\gamma_E(\ell)$ and $C^\gamma_B(\ell)$, the latter of which is zero if lensing on a scalar gravitational potential is considered and if the Born-approximation applies.

Figs.~\ref{fig_ellipticity_low} and~\ref{fig_ellipticity_high} shows intrinsic ellipticity spectra $C_E^\epsilon(\ell)$ and $C_B^\epsilon(\ell)$ for the EUCLID galaxy sample with its median redshift at $z_\mathrm{med}=0.9$ which are contrasted with the spectra $C_E^\epsilon(\ell)$ and $C_B^\epsilon(\ell)$ for a galaxy sample with a much lower median redshift of $z_\mathrm{med}=0.3$. For comparison, we superpose the corresponding spectra $C_E^\gamma(\ell)$ for the weak lensing shear $\gamma$ measured on the same galaxy populations, both for a linear and a nonlinear CDM-spectrum.

The spectra are constant and equal in amplitude up to multipoles of $\ell\simeq100$, indicating the absence of correlations such that on each scale on measures the variance of the uncorrelated ellipticity field. Correlations become important on angular scales $\ell\gsim 300$ where the spectra level off and decrease from multipoles of $\ell\gsim3000$ on very rapidly. In the peak region, the ellipticity $E$-modes have an amplitude larger than the $B$-modes by about an order of magnitude. Consistent with expectations the lensing spectrum exceeds intrinsic ellipticity correlations significantly for the high redshift case, but one observes the contrary behaviour for the low redshift case. Furthermore, the impact of nonlinear structure formation is stronger in the low-redshift case, as non-linearities had more time to develop.

\section{Imprints on lensing on ellipticities}\label{sect_evolution}

\subsection{Lensing effects on ellipticity fields}
In weak lensing studies it is commonly assumed that the observed lensed galaxies would have had no shape-correlations without lensing, and that there is no clustering in the galaxy sample, neither along the line-of-sight nor perpendicular to it. In short, intrinsic ellipticities are drawn independently from a distribution which is commonly assumed to be Gaussian with variance $\sigma_\epsilon^2$. In the estimation process of weak lensing spectra from galaxy shapes the intrinsic shape variations would contribute the Poissonian term $\sigma_\epsilon^2/n$, if the estimation process comprises $n$ galaxies.

The physical picture we have in mind is a background field of intrinsically correlated ellipticities on which gravitational lensing acts by deflection and shear, while there are no correlations between the ellipticities and the matter distribution responsible for lensing. In this sense, we consider lensing on II-alignments, while neglecting GI-alignments, which vanish in the case of Gaussian fluctuations statistics for quadratic alignment models as the one used in this work, but would be present in nonlinear structures \citep{0004-637X-681-2-798}. Other lensing-induced effects are modulations in the surface density of galaxies, due to the interplay between magnification of the image brightness and dilution over a larger solid angle, which we neglect for the purpose of this paper.

\subsection{Adaptation of the CMB-lensing formalism}
As in the case of CMB lensing where the (complex) polarisation tensor $P(\vect) = Q(\vect)+\ci U(\vect)$ is measured at a new position $\vect+\veca$ due to gravitational lensing, we assert that the ellipticity $\epsilon$ is not observed at the true position $\vect$ of the galaxy, but at the apparent position $\vect+\veca$, $\epsilon(\vect)\rightarrow\epsilon(\vect+\veca)$ with the lensing deflection angle $\veca$. Additionally, a variation of the deflection angle across the galaxy image leads to a distortion described by the complex shear $\gamma$ and the convergence $\kappa$.

Adapting the CMB-lensing formalism, correlations between the components of the shifting angle $\alpha$ at two positions $\vect_1$ and $\vect_2$ are described by \citep{1996ApJ...463....1S}
\begin{equation}
\bra\alpha_i(\vect_1)\alpha_j(\vect_2)\ket = \frac{1}{2}C_0(\theta) - C_2(\theta)\:\hat{\theta}_{\langle i}\hat{\theta}_{j\rangle}
\end{equation}
with $\vect=\vect_2-\vect_1$. The two correlation functions of the deflection angle are defined as
\begin{equation}
C_0(\theta) = \int\frac{\ell^3\dd\ell}{2\pi} C_\psi(\ell)J_0(\ell\theta)
\end{equation}
and
\begin{equation}
C_2(\theta) = \int\frac{\ell^3\dd\ell}{2\pi} C_\psi(\ell)J_2(\ell\theta).
\end{equation}
We abbreviate the variance of the deflection angle
\begin{equation}
\sigma^2(\theta) = C_0(0) - C_0(\theta)
\end{equation} 
in complete analogy to CMB-lensing for describing uncorrelated deflections. 

The characteristic function of $\veca$, i.e. the Fourier transform of the probability density $p(\veca)\dd\veca$ is then obtained as:
\begin{equation}
\big\langle\exp\left(\ci\bmath{\ell}\left[\veca(\vect_1)-\veca(\vect_2)\right]\right)\big\rangle = 
\exp\left(\frac{\ell^2}{2}\left[-\sigma^2(\theta) + \cos2\varphi_\ell C_2(\theta)\right]\right),
\end{equation}
and can be expressed in the case of Gaussian distributions in terms of $\sigma^2(\theta)$ and $C_2(\theta)$. In the case of CMB-lensing, non-Gaussian contributions have been shown to have negligible effect on the deflection angle statistics \citep{2009MNRAS.396..668C, 2011MNRAS.411.1067M} and in the case of weak cosmic shear, arguments about the rarity of strong deflections not described by a Gaussian distributions apply in a similar way \citep{2005MNRAS.356..829H}.

\begin{figure}
\begin{center}
\resizebox{\hsize}{!}{\includegraphics{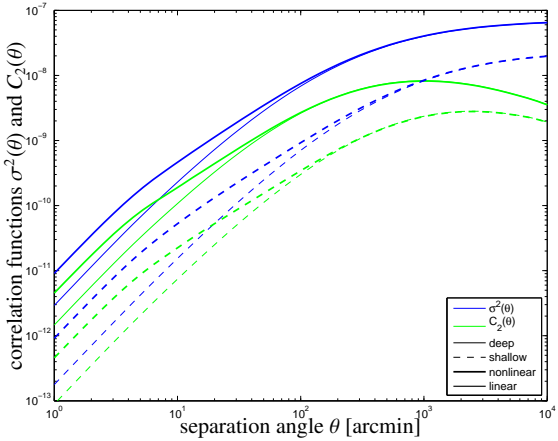}}
\end{center}
\caption{Correlation functions $\sigma^2(\theta) = C_0(0)-C_0(\theta)$ (blue line) and $C_2(\theta)$ (green line) as a function of separation angle $\theta$, for the high redshift galaxy sample (solid line) and the low redshift galaxy sample (dashed line), again comparing the predictions from linear (thin lines) and nonlinear (thick lines) CDM spectra.}
\label{fig_correlation}
\end{figure}

Fig.~\ref{fig_correlation} shows the quantities $\sigma^2(\theta) = C_0(0) - C_0(\theta)$ and $C_2(\theta)$ needed in this formalism, for both a high and a low redshift galaxy sample and for linear and nonlinear CDM-spectra: Both correlation functions assume larger values for the high-redshift sample and for lensing on nonlinear structures, which in particular causes a larger variance of the deflection angle on small angular scales. For large values of of the argument $\theta$, both correlation function start to oscillate rapidly.

The correlation properties of the lensing-distorted ellipticity field can be described using the two correlation functions $\xi_\pm(\theta)$,
\begin{eqnarray}
\xi_+^\prime(\theta) & = & 
\bra\epsilon^*(\vecx+\veca)\epsilon(\vecx^\prime+\veca^\prime)\ket\\
\xi_-^\prime(\theta) & = & 
\bra\exp(-4\ci\phi_\ell)\epsilon(\vecx+\veca)\epsilon(\vecx^\prime+\veca^\prime)\ket.
\end{eqnarray}
Substituting the correlation function for the deflection angle in the Fourier-transforms of the above expressions yields the correlation functions $\xi_\pm^\prime(\theta)$ of the new ellipticity field. They can be transformed to $E$-mode and $B$-mode spectra with the standard transformations eqns.~(\ref{eqn_e_transform}) and~(\ref{eqn_b_transform}).

These steps lead to a transformation formula the $E$-mode and $B$-mode spectra of the ellipticity field, which can be summarised by a concise matrix notation:
\begin{equation}
\left(
\begin{array}{c}
C_E^\prime(\ell) \\
C_B^\prime(\ell)
\end{array}
\right) = 
\int\lprime\dd\lprime
\left(
\begin{array}{cc}
W_+(\ell,\lprime) & W_-(\ell,\lprime) \\
W_-(\ell,\lprime) & W_+(\ell,\lprime)
\end{array}
\right)
\left(
\begin{array}{c}
C_E^\epsilon(\lprime)\\
C_B^\epsilon(\lprime)
\end{array}
\right).
\label{eqn_transformation}
\end{equation}
This notation shows explicitly the mixing between scales due to the convolution weighted with $W_+(\ell,\lprime)$ and the conversion between $C_E^\epsilon(\ell)$ and $C_B^\epsilon(\ell)$ under the influence of $W_-(\ell,\lprime)$. And clearly, the displacement mechanism can not generate ellipticity correlations as $C^\prime_X(\ell)$ remains zero if the $C^\epsilon_X(\ell)$ is zero to begin with, $X\in\left\{E,B\right\}$.

These kernels $W_\pm(\ell,\lprime)$ are given by
\begin{eqnarray}
W_+(\ell,\lprime) & = & \frac{1}{2}\int\theta\dd\theta \left[J_0(\ell\theta)A(\lprime,\theta) + J_4(\ell\theta)B(\lprime,\theta)\right],\\
W_-(\ell,\lprime) & = & \frac{1}{2}\int\theta\dd\theta \left[J_0(\ell\theta)A(\lprime,\theta) - J_4(\ell\theta)B(\lprime,\theta)\right],
\end{eqnarray}
with the functions
\begin{eqnarray}
A(\ell,\theta) & = & 
\exp\left(-\frac{\ell^2\sigma^2(\theta)}{2}\right)\left[J_0(\ell,\theta) + \frac{\ell^2}{2}C_2(\theta)J_4(\ell\theta)\right],\\
B(\ell,\theta) & = &
\exp\left(-\frac{\ell^2\sigma^2(\theta)}{2}\right)\left[J_4(\ell,\theta) + \frac{\ell^2}{2}C_2(\theta)J_s(\ell\theta)\right].
\end{eqnarray}
Here, uncorrelated deflections contained in the variance $\sigma^2(\theta)$ give rise to a Gaussian convolution kernel while correlated deflections due to $C_2(\theta)$ show a more complicated mode-coupling. We abbreviated $J_s(x) = J_2(x) + J_6(x)$. 

In the limit absent lensing, $C_0(\theta) = C_2(\theta) = 0$ such that $W_+(\ell,\lprime) = \delta(\ell-\lprime) / \ell$ and $W_-(\ell,\lprime) = 0$, due to the orthogonality relations of the cylindrical Bessel functions,
\begin{equation}
\int(\ell\theta)\:\dd\theta\: J_n(\ell\theta) J_n(\lprime\theta) = \delta_D(\ell-\lprime).
\end{equation}
In this case, the convolution is reduced to a Dirac $\delta_D$-function and the mixing matrix is the unit matrix, so that the $E$-mode and $B$-mode amplitudes are conserved and there is no convolution between $\ell$-modes. We have verified that higher-order corrections arising in the transformation of correlation functions do have a negligible effect for the evolved ellipticity correlations \citep{2005PhRvD..71j3010C, 2006PhR...429....1L} and in our numerical implementation, we used the same relations for the required number of grid points ($4\times\ell_\mathrm{max}$ tabulated values of $\sigma^2(\theta)$ and $C_2(\theta)$ in $\theta$) as suggested for CMB-lensing.

\subsection{Conversion between $E$ and $B$-modes}
\begin{figure}
\begin{center}
\resizebox{\hsize}{!}{\includegraphics{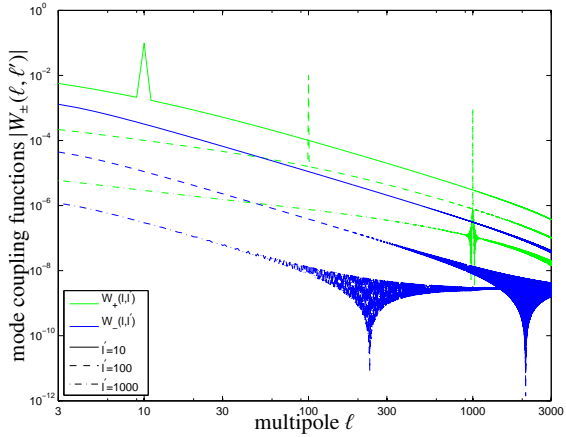}}
\end{center}
\caption{Mode coupling functions $W_+(\ell,\lprime)$ (green lines) and $W_-(\ell,\lprime)$ (blue lines) used in the transformation of the ellipticity spectra, which describes the convolving effect on the ellipticity spectra, for $\lprime=10,100,1000$ (solid, dashed and dash-dotted, respectively). The coupling functions shown are the ones for the high redshift galaxy sample, while those for the low redshift sample look qualitatively very similar.}
\label{fig_kernel}
\end{figure}

Fig.~\ref{fig_kernel} shows the mode coupling kernels $W_+(\ell,\lprime)$ and $W_-(\ell,\lprime)$ for the high-redshift distribution and computed for a nonlinear CDM spectrum, with qualitatively very similar results for the low redshift sample. Apart from a smooth variation of $W_+(\ell,\lprime)$, which acts on the ellipticity spectra by convolution, one notices tall spikes at $\ell=\lprime$, illustrating the closeness to diagonality of the $W_+$-matrix. In contrast, $W_-(\ell,\lprime)$ shows smaller amplitudes by about two orders of magnitude, indicating that the conversion between $E$- and $B$-modes is a minor effect compared to the convolution mediated by $W_+(\ell,\lprime)$, with strong oscillatory features close to diagonal $\ell=\lprime$. The kernels $W_\pm(\ell,\lprime)$ show an inverse scaling with multipole $\ell$ such that they become approximately constant when substituted into the relation~(\ref{eqn_transformation}) by multiplication with the $\lprime\dd\lprime$-differential.

\subsection{Lensed ellipticity spectra}
\begin{figure}
\begin{center}
\resizebox{\hsize}{!}{\includegraphics{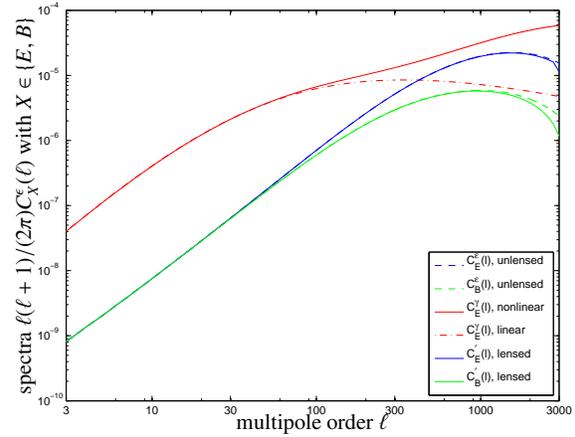}}
\end{center}
\caption{Ellipticity spectra $C_E^\epsilon(\ell)$ (blue line) and $C_B^\epsilon(\ell)$ (green line) as predicted by the angular momentum model (dashed lines), and the lensed ellipticity spectra (solid lines) where the deflections were computed for the high redshift galaxies. For comparison, we plot the spectrum $C_E^\gamma(\ell)$ of the weak lensing shear $\gamma$ (red line).}
\label{fig_ellipticity_low}
\end{figure}

\begin{figure}
\begin{center}
\resizebox{\hsize}{!}{\includegraphics{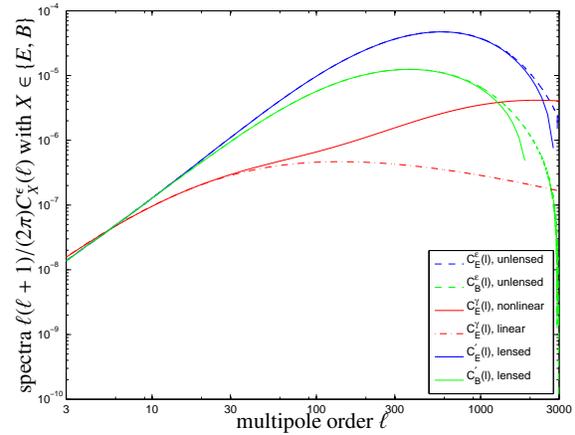}}
\end{center}
\caption{Ellipticity spectra $C_E^\epsilon(\ell)$ (blue line) and $C_B^\epsilon(\ell)$ (green line) as predicted by the angular momentum model (dashed lines), and the lensed ellipticity spectra (solid lines) where the deflections were computed for the low redshift galaxies. For comparison, we plot the spectrum $C_E^\gamma(\ell)$ of the weak lensing shear $\gamma$ for the same galaxy sample (red line).}
\label{fig_ellipticity_high}
\end{figure}

The final results are given in Figs.~\ref{fig_ellipticity_low} and~\ref{fig_ellipticity_high}, which compare the initial ellipticity spectra $C_E^\epsilon(\ell)$ and $C_B^\epsilon(\ell)$ of the ellipticity field as predicted by correlated angular momenta, and the distorted spectra $C_E^\prime(\ell)$ and $C_B^\prime(\ell)$ that encapsulate the imprint of lensing deflection and therefore display altered correlation properties. For comparison with weak lensing, we plot the weak convergence spectrum $C_\kappa(\ell)$ expected from the EUCLID galaxy sample in comparison, for a nonlinear CDM spectrum \citep[using the parametrisation by][]{2003MNRAS.341.1311S}. 

The first observation is that ellipticity correlations reach amplitudes similar to those of the weak lensing convergence in the nonlinear part corresponding to amplitudes $\ell\lsim300$, and that the intrinsic $E$-mode spectrum $C_E^\epsilon(\ell)$ is larger than the $B$-mode spectrum $C_B^\epsilon(\ell)$ by about an order of magnitude in this regime. On larger angular scales, there are no appreciable ellipticity correlations and one effectively observes the variance of the ellipticity field for uncorrelated objects. Consequently, the spectra have identical amplitudes and are effectively constant. In this regime, the shifting effect is not able to affect the galaxies, which is a well-known result in CMB-lensing, where scale free-spectra are invariant \citep{2006PhR...429....1L}: The mode-conversion mechanism is ineffective if the spectra are equal, $C_E^\epsilon(\ell) = C_B^\epsilon(\ell)$, and the convolution with $W_+(\ell,\lprime)$ is not able to redistribute amplitudes. In contrast, both spectra are affected on multipoles $\ell>1000$, where in particular $C_B^\prime(\ell)$ has decreased relative to $C_B^\epsilon(\ell)$.

\begin{figure}
\begin{center}
\resizebox{\hsize}{!}{\includegraphics{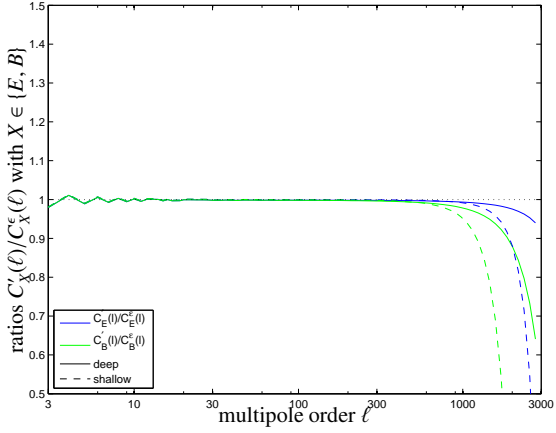}}
\end{center}
\caption{Ratios $C^\prime_E(\ell)/C^\epsilon_E(\ell)$ (blue lines) and $C^\prime_B(\ell)/C^\epsilon_B(\ell)$ (green lines), for both a high redshift galaxy sample (solid lines) and a low redshift galaxy sample (dashed lines).}
\label{fig_ratios}
\end{figure}

Fig.~\ref{fig_ratios} shows the changes in the spectra as a function of multipole $\ell$ by giving the ratio of the evolved and initial $E$-mode and $B$-mode spectra, $C_E^\prime(\ell)/C_E^\epsilon(\ell)$ and $C_B^\prime(\ell)/C_B^\epsilon(\ell)$ respectively. As already indicated by Figs.~\ref{fig_ellipticity_low} and~\ref{fig_ellipticity_high}, we see a significant decrease of amplitude amounting to 5\% for the $E$- and 30\% for the $B$-modes from $\ell = 3000$ on in the case of the high-redshift sample and from $\ell = 1000$ on in the case of the low-redshift sample. This implies that for EUCLID's weak lensing application, changes in the ellipticity spectra are affecting scales where the shape noise starts dominating, but for shallower surveys, lower multipoles would be affected by weak lensing deflection. We conclude that in the case of deep surveys such as EUCLID, weak lensing manifests itself primarily as weak lensing shear which dominates over intrinsic alignments and the lensing deflection effect shapes intrinsic alignments by decreasing their amplitudes only at very high multipoles.

\subsection{Violated homogeneity of the ellipticity field}
Lensing of the intrinsic ellipticity field introduces a violation of their statistical homogeneity in complete analogy to the lensing deflection acting on the CMB polarisation. For a given realisation of the deflection potential $\psi$ one can estimate the lensing effect on the ellipticity spectra, $X=E,B$:
\begin{equation}
\bra\epsilon^\prime_X(\bmath{\ell}) \epsilon^\prime_X(\bmath{\ell^\prime})\ket
=
f_X(\bmath{\ell},\bmath{\ell^\prime})
\psi(\bmath{\ell}-\bmath{\ell}^\prime)
\end{equation}
with 
\begin{eqnarray}
f_E(\bmath{\ell},\bmath{\ell^\prime}) & = & 
(\bmath{\ell} - \bmath{\ell}^\prime)
\left[\bmath{\ell} C_E^\epsilon(\ell) + \bmath{\ell}^\prime C_E^\epsilon(\ell^\prime)\right]
\cos2\varphi_{\ell,\ell^\prime} \\
f_B(\bmath{\ell},\bmath{\ell^\prime}) & = & 
(\bmath{\ell} - \bmath{\ell}^\prime)
\left[\bmath{\ell} C_B^\epsilon(\ell) + \bmath{\ell}^\prime C_B^\epsilon(\ell^\prime)\right]
\cos2\varphi_{\ell,\ell^\prime}
\end{eqnarray}
where $\varphi_{\ell,\ell^\prime}$ the enclosed angle. Nonzero correlations between multipoles are the signature of homogeneity violation introduced by a single realisation and they would disappear in the process of ensemble averaging the lensing potential.

In order to place an upper limit on this effect we select a particularly simple geometry, namely parallel alignment of the wave vectors $\bmath{\ell}$ and $\bmath{\ell}^\prime$ such that the cosines are equal to one and have the wave vectors them differ by one unit, $\ell - \lprime = 1$, as the coupling between neighbouring multipoles is strongest due to the rapid decline of the lensing potential's Fourier transform with increasing $\ell$. In this limit, and if one assumes the ellipticity spectra to be slowly varying, the ratio between the off-diagonal and diagonal correlations is given by
\begin{equation}
\frac{C_E^\prime(\ell,\ell+1)}{C_E^\epsilon(\ell)} =
\frac{C_B^\prime(\ell,\ell+1)}{C_B^\epsilon(\ell)} = 
\frac{\ell}{2\pi^2}\sigma_\psi(\ell = 1), \\
\end{equation}
where we replaced $\psi$ by $\sigma_\psi$ as an order of magnitude estimate for a typical amplitude. Substituting numbers yields upper limits of 10\% for the off-diagonal correlation relative to the diagonal ones at $\ell=10^3$ which is smaller than the weak lensing shear at high redshifts, but comparable if not slightly larger than the lensing shear at low redshifts.

As the imprints of weak lensing on the intrinsic ellipticity pattern are basically identical to those in the case of lensing of the CMB polarisation, it is conceivable that the lensing deflection field can be estimated by measuring the amount of off-diagonal (meaning $\ell\neq\lprime$) spectra of the ellipticity modes as illustrated above, by applying the reconstruction technique worked out by Hu and Okamoto: With a model for intrinsic ellipticity spectra (which can be well predicted using a good prior on $\Omega_m$, $\sigma_8$, $a$ and $\alpha$) the statistics of the lensing deflection field can be inferred from broken homogeneity. In contrast to polarisation tensors, however, the ellipticity field is strongly shaped by shear and only in extreme cases such as a low-redshift galaxy sample at large multipoles the intrinsic alignment effects are dominant, where of course issues with the low surface density of lensing galaxies and the corresponding high Poisson-noise become important.

\begin{figure}
\begin{center}
\resizebox{\hsize}{!}{\includegraphics{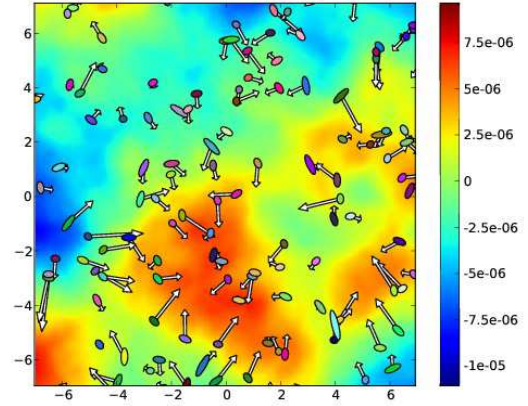}}
\end{center}
\caption{Realisation of a lensing potential on a $14^\circ\times14^\circ$-patch with 100 randomly sampled galaxies. The ellipticity is given by the lensing shear and the arrow indicates the apparent shift due to the weak lensing deflection. The realisation uses a lensing potential for the high-redshift galaxy sample with dominating shear. The displacement arrows are enhanced by a factor of 100 and the shears by a factor of 10, applied to a circular spot.}
\label{fig_illustration}
\end{figure}

For illustration, Fig.~\ref{fig_illustration} shows a realisation of the lensing potential with derived lensing deflections as gradients of the potential. The second derivatives have been used for generating a shear field that is depicted as a shape distortions of the otherwise circular spots. For visualisation, the deflections have been enlarged by a factor of 100 and the shear has been multiplied by 10.

\section{Summary}\label{sect_summary}
The subject of this work was an investigation of lensing effects and their observable signatures if the lensed galaxy sample shows intrinsic ellipticity correlations. Apart from weak lensing shear that operates on the shape of galaxies there will be a lensing deflection, which is unobservable in the case of uncorrelated ellipticities, but generates observable signatures if the lensing galaxies are intrinsically shape correlated. The lensing deflection manifests itself in the ellipticity spectra in three distinct ways: Firstly, there is a loss of amplitude in the spectra at high multipoles, secondly one observes a redistribution of amplitude between the $E$-mode and $B$-mode spectra, and thirdly there will be correlations between adjacent multipoles. In deriving these effects we made heavy use of analogies to the theory of lensing of the cosmic microwave background polarisation and identical mathematical properties of the ellipticity- the polarisation tensors. In our investigation, we are comparing the forecasts for a high-redshift lensing survey such as EUCLID with a low-redshift galaxy sample in order understand the scaling behaviour of all effects with distance.

\begin{enumerate}
\item{We derive ellipticity $E$-mode and $B$-mode spectra from a physical alignment model due to \citet{2001ApJ...559..552C}. This model is quadratic in the tidal shear and applicable for describing shape correlations between isolated spiral galaxies. Due to the lack of an analytical description of how a galactic disk is oriented inside a dark matter halo we think of our spectra as upper limits as we assume perfect alignment of the symmetry axis of the galactic disk with the host halo's angular momentum direction. The two parameters that enter our ellipticity model, the alignment parameter $a$ and the disk thickness $\alpha$, are determined from numerical simulations and from observations of local galaxies.}
\item{The impact of lensing deflection on intrinsically shape-correlated galaxies is threefold: There is a smoothing of the intrinsic ellipticity spectra, a mixing in multipole and a conversion between $E$-modes and $B$-modes, and the generation of correlations between otherwise uncorrelated multipoles, as an expression of violated homogeneity of the lensed galaxy field.}
\item{By drawing analogies between galaxy ellipticities and the CMB-polarisation, namely that both are tensorial fields with spin-2, we can formulate transformation formulas for the ellipticity spectra, if individual galaxies have been coherently shifted to a new position by lensing. The transformation formula can be written concisely as a combined convolution and mode-mixing relation.}
\item{Lensing deflection operates on intrinsic ellipticity spectra by convolution. Correlation amplitudes are redistributed in multipoles which can be observed on small angular scales when intrinsic alignment spectra cease to be constant and drop in amplitude. Then, lensing causes the spectra to drop faster. Qualitatively, these effects are weak at high redshifts and dominated by far by the weak shear signal, but are sizable at low redshifts, where the weak lensing shear is small. In this case, weak lensing can actually weaken shape-correlations by random redistribution of intrinsically aligned galaxies.}
\item{The losses in amplitude amount roughly to 5\% in $C_E^\epsilon(\ell)$ and to 30\% in $C_B^\epsilon(\ell)$ at $\ell=3000$ in the case of the high redshift sample and at $\ell=1000$ in the case of the low redshift sample. Compared to the convolution of the spectra the conversion between $E$- and $B$-modes is a minor effect.}
\item{We have derived an upper limit on the correlation between different multipoles due to broken homogeneity by using the fact that the correlations with between adjacent multipoles should be strongest. These correlations can be estimated to be at most $\sim10\%$ of the spectra at the largest multipoles, both for $E$- and $B$-modes, and are proportional to $\ell$.}
\item{Although we could take advantage of formal analogies between the CMB-polarisation and ellipticity fields, concerning symmetry properties, the description with spectra and the incorporation of the lensing effect we would like to emphasise that that in contrast to the CMB, lensing does not introduce a bispectrum into the ellipticity correlations. The CMB-lensing bispectrum is sourced by the integrated Sachs-Wolfe effect in the same potential that causes the lensing-deflection \citep{2000PhRvD..62d3007H}, and there is no analogous mechanism in the case of galaxy ellipticities.}
\end{enumerate}
We conclude that for deep-reaching lensing surveys intrinsic alignments are subdominant and that the shaping of their correlations by weak lensing deflection \citep[and by peculiar motion, which is of a similar order of magnitude, see][]{2012arXiv1202.1196G} is small compared to gravitational shear. At low redshifts, however, the situation is inverted: Intrinsic alignments dominate and the most important lensing effect is deflection. In contrast to the CMB-polarisation, it is doubtful if a violation of homogeneity of the ellipticity field introduced by lensing can be observed.

\section*{Acknowledgements}
Our work was supported by the German Research Foundation (DFG) within the framework of the excellence initiative through the Heidelberg Graduate School of Fundamental Physics. In particular, AGS acknowledges funding from the FRONTIER-programme, from the International Max Planck Research School for Cosmic Physics and from the Heidelberg Graduate School for Fundamental Physics. We would like to thank Philipp M. Merkel for his insight into CMB-lensing reconstructions and numerical advice, which he shared with us, and Vanessa M. B{\"o}hm for valuable comments.

\bibliography{bibtex/aamnem,bibtex/references}
\bibliographystyle{mn2e}

\appendix

\bsp

\label{lastpage}

\end{document}